\documentclass[sigconf]{acmart}

\renewcommand\footnotetextcopyrightpermission[1]{}
\usepackage{booktabs} 
\usepackage{flexisym}
\usepackage{bbm}
\usepackage{balance} 
\usepackage{multirow}
\usepackage{float}
\usepackage{figsize}
\usepackage{amsmath}
\usepackage{todonotes}

\copyrightyear{2019}
\acmYear{2019}
\setcopyright{acmcopyright}
\acmConference{RMSE@RecSys'19}{}{September 20, 2019, Copenhagen, Denmark} 
\acmPrice{15.00} 
\acmDOI{http******}
\acmISBN{ISBN *************}

\begin{document}
\fancyhead{}

\title{Multi-stakeholder Recommendation and its Connection to Multi-sided Fairness}

\author{Himan Abdollahpouri}
\affiliation{
 University of Colorado Boulder, USA}
 \email{himan.abdollahpouri@colorado.edu}
 \author{Robin Burke}
\affiliation{
 University of Colorado Boulder, USA}
 \email{robin.burke@colorado.edu}
\begin{abstract}

There is growing research interest in recommendation as a multi-stakeholder problem, one where the interests of multiple parties should be taken into account. This category subsumes some existing well-established areas of recommendation research including reciprocal and group recommendation, but a detailed taxonomy of different classes of multi-stakeholder recommender systems is still lacking. Fairness-aware recommendation has also grown as a research area, but its close connection with multi-stakeholder recommendation is not always recognized. In this paper, we define the most commonly observed classes of multi-stakeholder recommender systems and discuss how different fairness concerns may come into play in such systems. 

\end{abstract}


%
%
\maketitle

\section{Introduction}
{\let\thefootnote\relax\footnote{RMSE workshop at ACM RecSys 2019, September 20, Copenhagen, Denmark.}}

Recommender systems (RS) have been used in a variety of different domains to help users find relevant and interesting items. They recommend movies to watch \cite{lekakos2008hybrid} , songs to listen \cite{celma2010music} , jobs to take \cite{paparrizos2011machine} or even a person to date \cite{pizzato2010recon}.

The research in RS has been mainly focused on the personalization. That is, delivering the recommendations that best match the needs and interests of the end user. That is indeed an important consideration as users are one of the most important stakeholders in any recommendation platform but not the only one \cite{abdollahpouri2019beyond}.
There are numerous examples of recommender systems in which there are other stakeholders that their needs and preferences should be taken into account. The incorporation of the objectives and preferences of different stakeholders in the recommendation process is referred to as \textit{multi-stakeholder recommendation} \cite{DBLP:conf/um/BurkeAMG16} 

Multi-stakeholder recommendation is a relatively new topic (at least in academia) and, therefore, there is still no clear understanding of what type of recommender system is a multi-stakeholder one. Defining different classes of multi-stakeholder recommendation would help to distinguish among different recommendation problems ,which is important for developing the right algorithmic solutions for these types of recommender systems. 
Moreover, fairness-aware recommendation as a growing sub-topic in recommendation research has gained a lot of attention in recent years. Fairness in recommendation can be defined in different ways, as discussed in \cite{yao2017beyond}. One of its accepted definitions is to avoid giving discriminatory recommendations to different users based on some sensitive features such as gender or race \cite{kamishima2012enhancement}. Talking about the fairness of a recommendation can be meaningful only when there are different stakeholders involved. Thus, there is a close connection between  the requirements of multi-stakeholder recommendation and those of fairness-aware recommendation.



In this paper, we present three most commonly observed classes of multi-stakeholder recommendation:
\begin{itemize}
    \item {Multi-receiver recommendation}
    \item {Multi-provider recommendation}
    \item {Recommendation with side stakeholders}
\end{itemize}

In addition, we discuss the fairness concerns that could come into play in these types of recommendation systems with regard to different stakeholders such as users, item providers, etc. 

\section{Multi-stakeholder recommendation}

\begin{figure}
    \centering
    \includegraphics[width=3.2in]{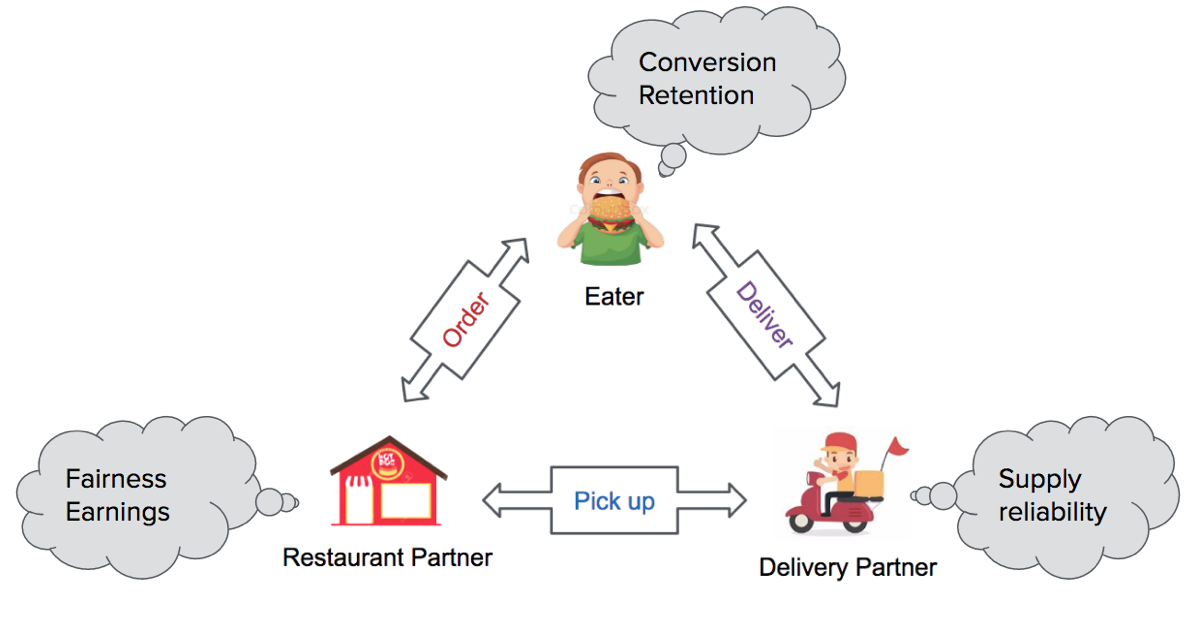}
    \caption []{A multi-stakeholder problem at Uber Eats \footnote{https://eng.uber.com/uber-eats-recommending-marketplace/}}
    \label{fig:uber}
\end{figure}
\footnotetext{https://eng.uber.com/uber-eats-recommending-marketplace/}

Recommender systems are often multi-stakeholder environments \cite{DBLP:conf/um/BurkeAMG16,umapHimanMS}: there are several stakeholders (often with conflicting preferences) whose needs and preferences should be taken into account in generating the recommendations. For example, in a food delivery system like the one for Uber Eats, we can observe three major stakeholders as shown in Figure ~\ref{fig:uber}: 1) the eaters or otherwise users who use the app and receive the recommendations for different restaurants, 2) the restaurant that are being recommended and 3) the delivery partners that take the food from the restaurants to the user's address. Eaters want to get relevant and interesting recommendations; restaurants want to be given a fair exposure to different users so they can have enough customers and finally the delivery partners need to be satisfied with the type of orders they need to deliver. Uber Eats cannot survive without the existence of any of these three parties and therefore it needs to take all of these stakeholders' preferences into account. This example is clearly beyond the standard user-focused recommender system in which the needs and preferences of the end user is the only consideration. 


Although the research in recommender system has been mainly focused on the satisfaction of the end user, there exist several threads of research in the recommender system that are  examples of multi-stakeholder recommendation but similarities among them have not typically been recognized. For example, in a group recommendation platform \cite{felfernig2018group} the system wants to recommend an item or a list of items to a group of users such that it meets the preferences of the group members. That is different from recommending items to only one user as, in many cases, the preferences of the users in the group could conflict with each other and, therefore, it is important for the recommender system to find ways to handle such conflicts. 

Another area of work that also falls into the multi-stakeholder definition is the reciprocal recommendation where the preferences of both the receiver and the provider of the recommendations should be taken into account \cite{yu2011reciprocal}. For example in online dating applications \cite{reciprocal,reciprocaldating}, recommending a user to another user is only meaningful if both users are happy with this recommendation.

Multi-stakeholder recommendation can be seen as an umbrella for many different types of recommendation problems involving two or more different stakeholders. These types of systems have not yet been studied in a systematic way and there is still a lot of room for further research and development. 

In next section, we define the most commonly observed classes of multi-stakeholder recommendation and provide examples for each. We believe having a well-defined architecture for these systems can facilitate further research in this area. 

\section{Classes of Multi-stakeholder recommendation}
In this section we define different classes of multi-stakeholder recommendation according to the architecture of the recommendation platform. One could argue there exist more types of architectures but we believe these are typically the main and commonly observed patterns.

\subsection{Multi-receiver Recommendation}
\begin{figure*}[t]
\centering
\SetFigLayout{2}{1}
  \subfigure[Heterogeneous multi-receiver recommendation]{\fbox{\includegraphics[height=2.2in]{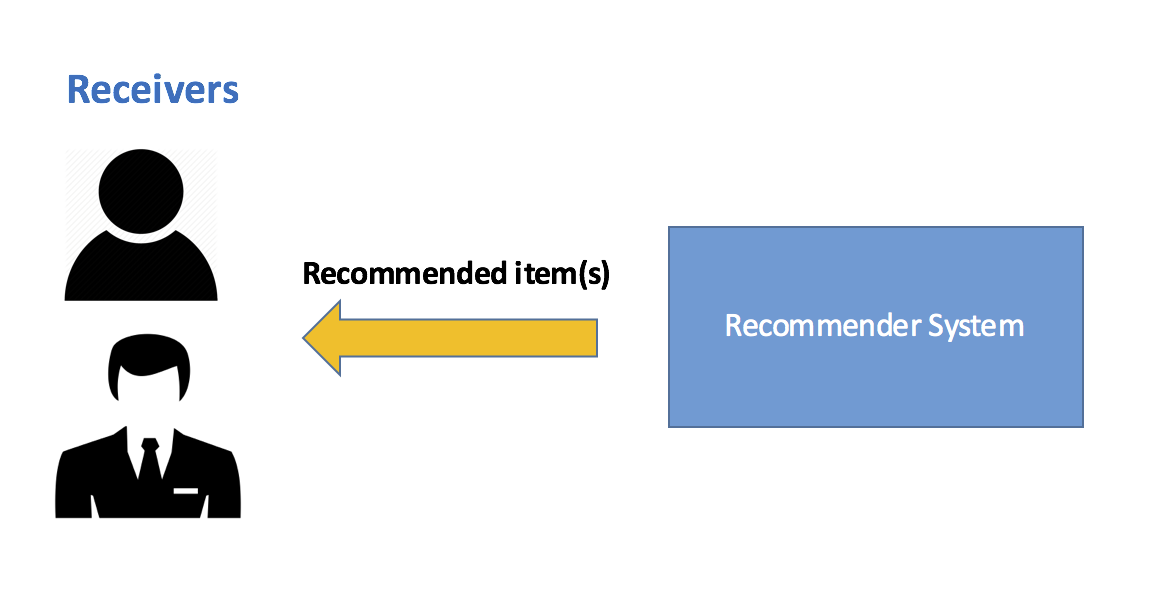}}}
  \hfill
  \subfigure[Homogeneous multi-receiver recommendation]{\fbox{\includegraphics[height=2in]{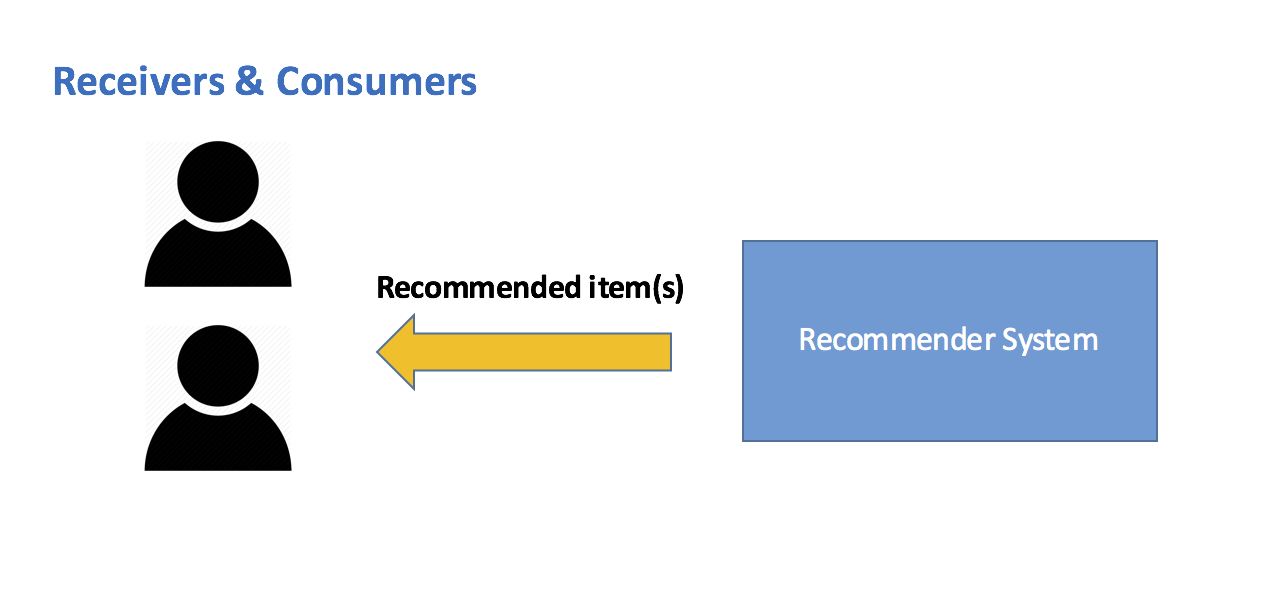}}}
  \hfill
\caption{The architectures for a) Heterogeneous and b) Homogeneous multi-receiver recommendations }
\end{figure*}\label{multi_receiver}

The first class of multi-stakeholder recommendation is \textit{Multi-receiver Recommendation} where the receiver of the recommendation is not one individual but rather a group of individuals. In these systems, the recommendation should be appealing to all the individuals receiving the recommendation. Depending on the characteristics of the receivers of the recommendation, there could exist two types of multi-receiver recommendation:

\begin{itemize}
    \item \textbf{Heterogeneous}:
    In this type of multi-receiver recommendation, the receivers of the recommendation may not necessarily be the same type. For example, in an educational system that recommends courses to students, the receivers of the recommendations are the students who take those courses. In addition, in many cases, the parents of those students could be also considered as the receiver of the recommendation even though they are not taking the course themselves but they somehow involve in the challenges and requirements that comes with those recommended course (e.g. paying the fee for the registration etc.) so their preferences should be also taken into account in recommending those courses. Zheng et al. \cite{zheng2019personalized} proposed a utility-based multi-stakeholder recommendations to the area of personalized learning in educations in which the preferences of students and instructors are taken into account simultaneously. Figure 2-a shows a typical architecture for this type of systems. We showed the users with different icons to clarify the heterogeneousness of the receivers. 
    \item \textbf{Homogeneous}: In contrast to the heterogeneous multi-receiver recommendation, users in homogeneous multi-receiver recommendation are all from the same type (e.g. students) and they are consuming the recommended item. This type of multi-receiver recommendation is known as \textit{group recommendation} in the literature where an item or set of items is recommended to a group of users who all use or consume the recommendation. For example, a movie recommender system that recommends movies to a group of friends or family members falls within this category as all these users are the real consumers of these recommendations. See the survey in \cite{masthoff2011group}. Figure 2-b  shows a typical architecture for this type of systems. The users are all shown with the same icon. 
\end{itemize}

\subsection{Multi-provider Recommendation}

\begin{figure*}
    \centering
    \fbox{\includegraphics[height=2.3in]{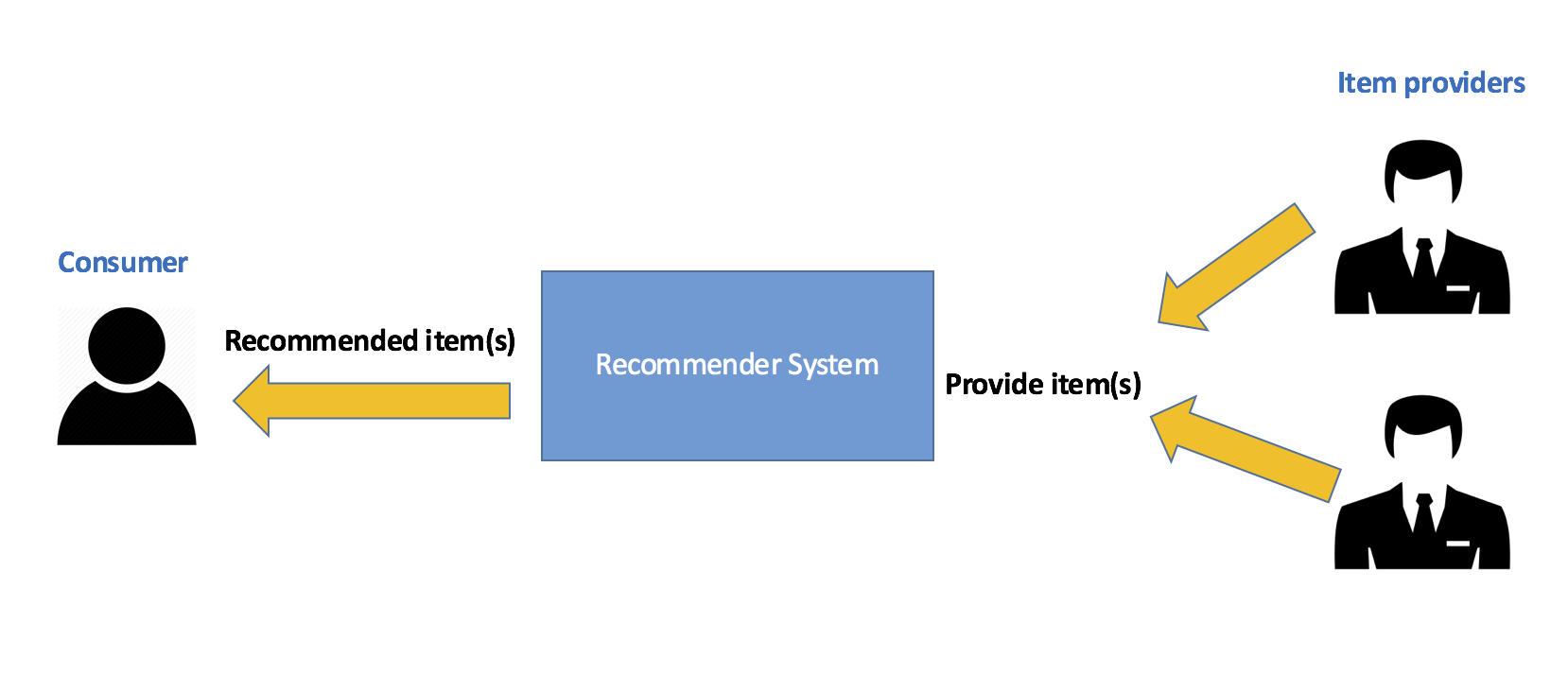}}
    \caption{Multi-provider recommender system}
    \label{fig:multi-provider}
\end{figure*}

It is often the case that the items and products on a recommender system are provided by several parties that use the platform for reaching out to their desired audience. For example, in a sharing economy platform like Airbnb all the available listings on the platform are actually provided by different hosts and the Airbnb itself does not own any of these apartments. These hosts are using the Airbnb platform to find travellers that could be interested in staying at their room or apartment. Therefore, Airbnb is a \textit{multi-provider recommendation} platform since there are numerous number of hosts in the system and Airbnb should try to give a fair amount of exposure to the listings provided by each of these hosts. Several interesting challenges could arise form this type of systems such as cold start providers (a provider that is new to the system and needs to get attention), malicious providers (a provider who is violating some rules or s/he has bad ratings from the users) and problems related to fairness which we discuss later in section ~\ref{fairness}.  Figure ~\ref{fig:multi-provider} shows the architecture for this type of systems. As you can see, multiple providers are giving their items to the system to be recommended to the users. The multi-provider recommender can be seen in two different types depending on whether or not the providers have preferences towards what type of users they want to reach out to:

\begin{figure*}
    \centering
   \fbox{\includegraphics[height=2.5in]{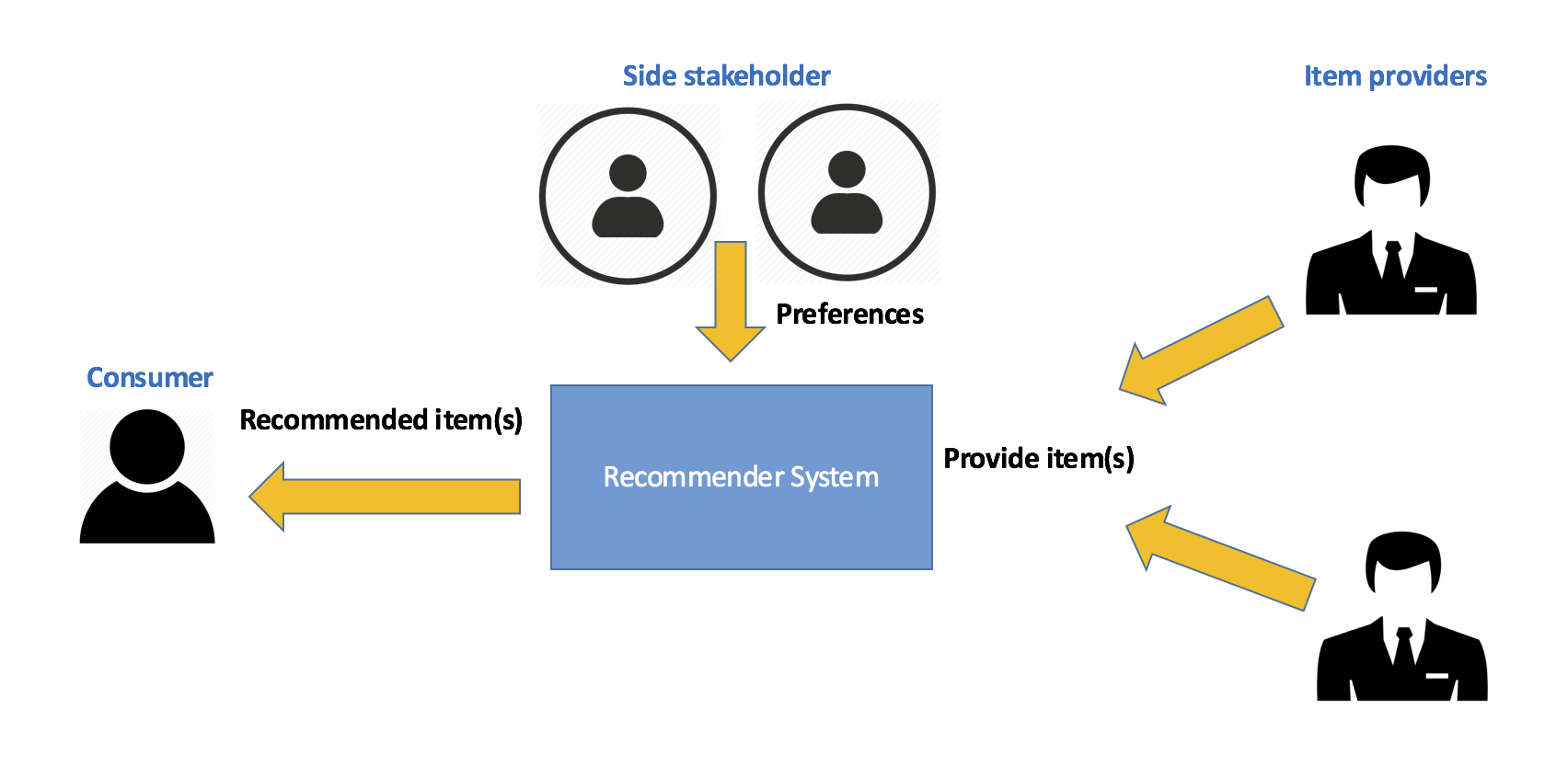}}
    \caption{Recommendation with side stakeholders}
    \label{fig:side-stakeholders}
\end{figure*}

\begin{itemize}
    \item \textbf{Without Provider Preferences}
    In this type of multi-provider recommendation the system is aware that there are multiple providers behind the items and they should get a fair exposure in the recommendations. However, the providers themselves have no preferences towards certain users to whom their items want to be recommended. For example, on the Kiva.org microlending platform, the items which are recommended are loans. That means the providers in this case are the people who asked for loan on the platform (i.e. the borrowers). In this example, the borrowers do not really care to whom their loan requests are recommended as long as the loan has a good chance of being funded. In other words, they do not have a specific target audience in mind.
     \item \textbf{With Provider Preferences}
     The providers in this case actually have certain type of audience in mind and the recommender system tries to fulfill the providers' preferences. For example, in \textit{computational advertising} an ad should be recommended to a user if s/he falls within the predefined groups of users who should get that particular ad~\cite{zhang2014optimal}. In other words, in addition to the ad not being annoying to the user (sadly it is often the case), the user should also be acceptable from the advertiser's perspective as they might want to reach certain users with a specific characteristics to maximize their ad efficiency.  
\end{itemize}
It is worth mentioning that, in some literature, the multi-provider recommendation is referred to as two-sided market \cite{mehrotra2018towards} where on one side we have the consumers and on the other side we have the item providers. We believe the name multi-provider recommendation can better explain the behavior of these type of recommendation systems as it clearly emphasizes the other side of the recommendation (besides the consumers): the item providers. 

\subsection{Recommendation with Side Stakeholders}
Stakeholders in a recommender system may not be always a direct part of the recommendation interaction. In other words they do not have to necessarily be the consumer of recommendations nor do they have to be the provider of the items being recommended. In some recommendation problems, there are other parties that are being affected by a given recommendation. In such cases, the satisfaction of these \textit{side stakeholders} should be taken into account in recommending items to users. For instance, on Uber Eats, when a list of restaurants is being recommended to a user, the person who is responsible for delivering food to the user is also potentially affected by this recommendation. Not only are the driver's preferences for a given recommendation important, the overall economy of the amount of work load given to the delivery partners is also something that the system should consider. 

The main characteristics of this type of recommendation systems is the fact that other stakeholders who are affected by the recommendations are not the entities who provide the recommendations (multi-provider) nor are they the users who receive the recommendations (multi-receiver). As in the Uber Eats example, the delivery partners are not receiving food recommendations nor are they providing the recommendations (like restaurants) but they still are affected by these recommendations. Figure ~\ref{fig:side-stakeholders} shows the architecture of a recommendation platform with side stakeholders. In this model, there are some side stakeholders in addition to the receivers and item providers whose preferences should be taken into account by the recommender system. One interesting challenge in this type of systems is how the system should aggregate those preferences and to what extent each stakeholder's preferences should be incorporated or prioritized in this decision making process. 

\subsubsection{Value-aware recommendation}
A frequent instantiation of a recommendation with side stakeholders is known as value-aware recommendation \cite{burke2017vams,pei2019value} where the recommendation platform is also considered as a stakeholder as it may have some goals and preferences with respect to the recommendations--considerations such as profit maximization \cite{azaria2013movie,JannachAdomaviciusVAMS2017}, long-tail promotion \cite{szpektor2011improving,flairs2019} are examples of such goals and preferences. These systems are called \textit{value-aware recommendation} because the recommender system needs to make sure, in addition to the users' satisfaction, the recommendations bring some sort of value to the business. For example, authors in \cite{JannachAdomaviciusVAMS2017} proposed a value-aware movie recommendation that takes both the price for each movie and also the user satisfaction into account in generating the recommendations. They showed it was possible to significantly increase the revenue with a negligible loss in accuracy.  

\subsection{Hybrid}
These were the most commonly observed classes of multi-stakeholder recommendation. One can imagine a combination of any of these classes. For example, the Uber Eats platform is a combination of multi-provider recommender system (different restaurants) and recommender with side stakeholders (the delivery partners). In fact, Uber Eats could be also a value-aware recommender system as the Uber Eats itself might also have certain business goals in mind. 

\section{Fairness in Multi-stakeholder
recommendation}\label{fairness}
Recently, fairness-aware recommendation has attracted a lot of attention from the recommender systems research community \cite{yao2017beyond,burke2017balanced,lee_fairness-aware_2014}. As we discussed earlier, there is a very close connection between multi-stakeholder recommendation and fairness-aware recommendation as having multiple stakeholders gives rise to the questions of fairness in recommendation. As in \cite{abdollahpouri2019beyond} we can see different classes
of fairness, distinguished by the fairness issues that
arise relative to what authors call different sides of the platform: consumers (C-fairness) and 
providers (P-fairness). In this paper, as we saw in section 3.4, there could be some stakeholders that are neither the receiver nor are they the provider of the recommendations. Therefore, in addition to the mentioned fairness types, we define another type which accounts for the fairness towards other affected stakeholders in the system. We call this type of fairness \textit{S-Fairness} which refers to the idea of the recommender system being fair to the \textit{side} stakeholders who are not directly participating in the recommendation process but rather being affected by such transaction. Therefore, in a multi-stakeholder recommendation platform we can observe the following types of fairness:

\begin{itemize}
    \item \textbf{C-fairness}: A recommender system distinguished by C-fairness is one that must take into account the disparate impact of recommendation on protected classes of recommendation consumers. For example, in a job recommender system that has a C-fairness requirement, the recommendations should be fair towards the users in the protected class (as defined by gender, age, nationality, etc.) relative to other users.
    
    In this paper, however, we extend the \textit{C-fairness} to any fairness concerns that the system might have with respect to the consumers of the recommendation. For instance, in a group recommender system the \textit{C-fairness} could refer to taking into account the preferences of different members of the group in a fair way and avoid ignoring certain users' preferences for a long term \cite{xiao2017fairness}. 
    \item \textbf{P-fairness}: A recommender system that has a P-fairness requirement should treat the providers of the items in a fair way. For example, in job recommendation platform, it could mean that minority-owned businesses have their jobs recommended to qualified candidates. 
    \item \textbf{S-Fairness} There are examples of multi-stakeholder recommendation in which some stakeholder are not the receiver nor are they the providers of the recommendations. Instead, they act as \textit{side} stakeholders whose needs and preferences should be taken into account in a fair way. For example, in the Uber Eats example, the delivery partners are side stakeholders and, therefore, the system needs to generate the recommendations in a way that the fairness considerations towards these delivery partners such as fair work load, fair commute distance etc. are being addressed properly.  
\end{itemize}

In particular applications, multiple fairness concerns may arise on different sides of the interaction. Thus, a system may have any combination of these fairness considerations in play at once: CP-fairness, for both consumers and providers, as noted in \cite{abdollahpouri2019beyond}, but also any of the others.

\section{Conclusion and Future work}
In this paper we defined several most commonly observed classes of multi-stakeholder recommendation. We argued how some of the already existing recommendation problems such as group recommendation and reciprocal recommendation are also examples of multi-stakeholder recommendation. In addition, we discussed several different types of fairness that are important to address in a multi-stakeholder recommendation. For future work, we intend to develop algorithmic solutions for each of these different types of multi-stakeholder recommendation. 

\bibliographystyle{ACM-Reference-Format}
\bibliography{main.bib}


\begin{thebibliography}{26}


\ifx \showCODEN    \undefined \def \showCODEN     #1{\unskip}     \fi
\ifx \showDOI      \undefined \def \showDOI       #1{#1}\fi
\ifx \showISBNx    \undefined \def \showISBNx     #1{\unskip}     \fi
\ifx \showISBNxiii \undefined \def \showISBNxiii  #1{\unskip}     \fi
\ifx \showISSN     \undefined \def \showISSN      #1{\unskip}     \fi
\ifx \showLCCN     \undefined \def \showLCCN      #1{\unskip}     \fi
\ifx \shownote     \undefined \def \shownote      #1{#1}          \fi
\ifx \showarticletitle \undefined \def \showarticletitle #1{#1}   \fi
\ifx \showURL      \undefined \def \showURL       {\relax}        \fi
\providecommand\bibfield[2]{#2}
\providecommand\bibinfo[2]{#2}
\providecommand\natexlab[1]{#1}
\providecommand\showeprint[2][]{arXiv:#2}

\bibitem[\protect\citeauthoryear{Abdollahpouri, Adomavicius, Burke, Guy,
  Jannach, Kamishima, Krasnodebski, and Pizzato}{Abdollahpouri
  et~al\mbox{.}}{2019a}]%
        {abdollahpouri2019beyond}
\bibfield{author}{\bibinfo{person}{Himan Abdollahpouri},
  \bibinfo{person}{Gediminas Adomavicius}, \bibinfo{person}{Robin Burke},
  \bibinfo{person}{Ido Guy}, \bibinfo{person}{Dietmar Jannach},
  \bibinfo{person}{Toshihiro Kamishima}, \bibinfo{person}{Jan Krasnodebski},
  {and} \bibinfo{person}{Luiz Pizzato}.} \bibinfo{year}{2019}\natexlab{a}.
\newblock \showarticletitle{Beyond Personalization: Research Directions in
  Multistakeholder Recommendation}.
\newblock \bibinfo{journal}{\emph{arXiv preprint arXiv:1905.01986}}
  (\bibinfo{year}{2019}).
\newblock


\bibitem[\protect\citeauthoryear{Abdollahpouri, Burke, and
  Mobasher}{Abdollahpouri et~al\mbox{.}}{2017}]%
        {umapHimanMS}
\bibfield{author}{\bibinfo{person}{Himan Abdollahpouri}, \bibinfo{person}{Robin
  Burke}, {and} \bibinfo{person}{Bamshad Mobasher}.}
  \bibinfo{year}{2017}\natexlab{}.
\newblock \showarticletitle{Recommender systems as multi-stakeholder
  environments}. In \bibinfo{booktitle}{\emph{Proceedings of the 25th
  Conference on User Modeling, Adaptation and Personalization (UMAP2017)}}.
  ACM.
\newblock


\bibitem[\protect\citeauthoryear{Abdollahpouri, Burke, and
  Mobasher}{Abdollahpouri et~al\mbox{.}}{2019b}]%
        {flairs2019}
\bibfield{author}{\bibinfo{person}{Himan Abdollahpouri}, \bibinfo{person}{Robin
  Burke}, {and} \bibinfo{person}{Bamshad Mobasher}.}
  \bibinfo{year}{2019}\natexlab{b}.
\newblock \showarticletitle{Managing Popularity Bias in Recommender Systems
  with Personalized Re-ranking.}. In \bibinfo{booktitle}{\emph{Florida AI
  Research Symposium (FLAIRS)}}. \bibinfo{publisher}{ACM}, \bibinfo{pages}{To
  appear}.
\newblock


\bibitem[\protect\citeauthoryear{Azaria, Hassidim, Kraus, Eshkol, Weintraub,
  and Netanely}{Azaria et~al\mbox{.}}{2013}]%
        {azaria2013movie}
\bibfield{author}{\bibinfo{person}{Amos Azaria}, \bibinfo{person}{Avinatan
  Hassidim}, \bibinfo{person}{Sarit Kraus}, \bibinfo{person}{Adi Eshkol},
  \bibinfo{person}{Ofer Weintraub}, {and} \bibinfo{person}{Irit Netanely}.}
  \bibinfo{year}{2013}\natexlab{}.
\newblock \showarticletitle{Movie recommender system for profit maximization}.
  In \bibinfo{booktitle}{\emph{Proceedings of the 7th ACM conference on
  Recommender systems}}. ACM, \bibinfo{pages}{121--128}.
\newblock


\bibitem[\protect\citeauthoryear{Burke, Adomavicius, Guy, Krasnodebski,
  Pizzato, Zhang, and Abdollahpouri}{Burke et~al\mbox{.}}{2017a}]%
        {burke2017vams}
\bibfield{author}{\bibinfo{person}{Robin Burke}, \bibinfo{person}{Gediminas
  Adomavicius}, \bibinfo{person}{Ido Guy}, \bibinfo{person}{Jan Krasnodebski},
  \bibinfo{person}{Luiz Pizzato}, \bibinfo{person}{Yi Zhang}, {and}
  \bibinfo{person}{Himan Abdollahpouri}.} \bibinfo{year}{2017}\natexlab{a}.
\newblock \showarticletitle{VAMS 2017: Workshop on Value-Aware and
  Multistakeholder Recommendation}. In \bibinfo{booktitle}{\emph{Proceedings of
  the Eleventh ACM Conference on Recommender Systems}}. ACM,
  \bibinfo{pages}{378--379}.
\newblock


\bibitem[\protect\citeauthoryear{Burke, Sonboli, Mansoury, and
  Ordo{\~n}ez-Gauger}{Burke et~al\mbox{.}}{2017b}]%
        {burke2017balanced}
\bibfield{author}{\bibinfo{person}{Robin Burke}, \bibinfo{person}{Nasim
  Sonboli}, \bibinfo{person}{Masoud Mansoury}, {and} \bibinfo{person}{Aldo
  Ordo{\~n}ez-Gauger}.} \bibinfo{year}{2017}\natexlab{b}.
\newblock \showarticletitle{Balanced Neighborhoods for Fairness-aware
  Collaborative Recommendation}. In \bibinfo{booktitle}{\emph{Workshop on
  Responsible Recommendation (FATRec)}}.
\newblock


\bibitem[\protect\citeauthoryear{Burke, Abdollahpouri, Mobasher, and
  Gupta}{Burke et~al\mbox{.}}{2016}]%
        {DBLP:conf/um/BurkeAMG16}
\bibfield{author}{\bibinfo{person}{Robin~D. Burke}, \bibinfo{person}{Himan
  Abdollahpouri}, \bibinfo{person}{Bamshad Mobasher}, {and}
  \bibinfo{person}{Trinadh Gupta}.} \bibinfo{year}{2016}\natexlab{}.
\newblock \showarticletitle{Towards Multi-Stakeholder Utility Evaluation of
  Recommender Systems}. In \bibinfo{booktitle}{\emph{Workshop on Surprise,
  Opposition, and Obstruction in Adaptive and Personalized Systems, UMAP
  2016}}.
\newblock


\bibitem[\protect\citeauthoryear{Celma}{Celma}{2010}]%
        {celma2010music}
\bibfield{author}{\bibinfo{person}{Oscar Celma}.}
  \bibinfo{year}{2010}\natexlab{}.
\newblock \showarticletitle{Music recommendation}.
\newblock In \bibinfo{booktitle}{\emph{Music recommendation and discovery}}.
  \bibinfo{publisher}{Springer}, \bibinfo{pages}{43--85}.
\newblock


\bibitem[\protect\citeauthoryear{Felfernig, Boratto, Stettinger, and
  Tkal{\v{c}}i{\v{c}}}{Felfernig et~al\mbox{.}}{2018}]%
        {felfernig2018group}
\bibfield{author}{\bibinfo{person}{Alexander Felfernig},
  \bibinfo{person}{Ludovico Boratto}, \bibinfo{person}{Martin Stettinger},
  {and} \bibinfo{person}{Marko Tkal{\v{c}}i{\v{c}}}.}
  \bibinfo{year}{2018}\natexlab{}.
\newblock \bibinfo{booktitle}{\emph{Group recommender systems: An
  introduction}}.
\newblock \bibinfo{publisher}{Springer}.
\newblock


\bibitem[\protect\citeauthoryear{Jannach and Adomavicius}{Jannach and
  Adomavicius}{2017}]%
        {JannachAdomaviciusVAMS2017}
\bibfield{author}{\bibinfo{person}{Dietmar Jannach} {and}
  \bibinfo{person}{Gediminas Adomavicius}.} \bibinfo{year}{2017}\natexlab{}.
\newblock \showarticletitle{Price and Profit Awareness in Recommender Systems}.
  In \bibinfo{booktitle}{\emph{Proceedings of the ACM RecSys 2017 Workshop on
  Value-Aware and Multi-Stakeholder Recommendation}}. \bibinfo{address}{Como,
  Italy}.
\newblock


\bibitem[\protect\citeauthoryear{Kamishima, Akaho, Asoh, and Sakuma}{Kamishima
  et~al\mbox{.}}{2012}]%
        {kamishima2012enhancement}
\bibfield{author}{\bibinfo{person}{Toshihiro Kamishima},
  \bibinfo{person}{Shotaro Akaho}, \bibinfo{person}{Hideki Asoh}, {and}
  \bibinfo{person}{Jun Sakuma}.} \bibinfo{year}{2012}\natexlab{}.
\newblock \showarticletitle{Enhancement of the Neutrality in Recommendation.}.
  In \bibinfo{booktitle}{\emph{Decisions@ RecSys}}. \bibinfo{pages}{8--14}.
\newblock


\bibitem[\protect\citeauthoryear{Lee, Lou, Chen, Chen, Lin, Chiang, and
  Chen}{Lee et~al\mbox{.}}{2014}]%
        {lee_fairness-aware_2014}
\bibfield{author}{\bibinfo{person}{Eric~L. Lee}, \bibinfo{person}{Jing-Kai
  Lou}, \bibinfo{person}{Wei-Ming Chen}, \bibinfo{person}{Yen-Chi Chen},
  \bibinfo{person}{Shou-De Lin}, \bibinfo{person}{Yen-Sheng Chiang}, {and}
  \bibinfo{person}{Kuan-Ta Chen}.} \bibinfo{year}{2014}\natexlab{}.
\newblock \showarticletitle{Fairness-{Aware} {Loan} {Recommendation} for
  {Microfinance} {Services}}. \bibinfo{publisher}{ACM Press},
  \bibinfo{pages}{1--4}.
\newblock


\bibitem[\protect\citeauthoryear{Lekakos and Caravelas}{Lekakos and
  Caravelas}{2008}]%
        {lekakos2008hybrid}
\bibfield{author}{\bibinfo{person}{George Lekakos} {and}
  \bibinfo{person}{Petros Caravelas}.} \bibinfo{year}{2008}\natexlab{}.
\newblock \showarticletitle{A hybrid approach for movie recommendation}.
\newblock \bibinfo{journal}{\emph{Multimedia tools and applications}}
  \bibinfo{volume}{36}, \bibinfo{number}{1-2} (\bibinfo{year}{2008}),
  \bibinfo{pages}{55--70}.
\newblock


\bibitem[\protect\citeauthoryear{Masthoff}{Masthoff}{2011}]%
        {masthoff2011group}
\bibfield{author}{\bibinfo{person}{Judith Masthoff}.}
  \bibinfo{year}{2011}\natexlab{}.
\newblock \showarticletitle{Group recommender systems: Combining individual
  models}.
\newblock In \bibinfo{booktitle}{\emph{Recommender systems handbook}}.
  \bibinfo{publisher}{Springer}, \bibinfo{pages}{677--702}.
\newblock


\bibitem[\protect\citeauthoryear{Mehrotra, McInerney, Bouchard, Lalmas, and
  Diaz}{Mehrotra et~al\mbox{.}}{2018}]%
        {mehrotra2018towards}
\bibfield{author}{\bibinfo{person}{Rishabh Mehrotra}, \bibinfo{person}{James
  McInerney}, \bibinfo{person}{Hugues Bouchard}, \bibinfo{person}{Mounia
  Lalmas}, {and} \bibinfo{person}{Fernando Diaz}.}
  \bibinfo{year}{2018}\natexlab{}.
\newblock \showarticletitle{Towards a fair marketplace: Counterfactual
  evaluation of the trade-off between relevance, fairness \& satisfaction in
  recommendation systems}. In \bibinfo{booktitle}{\emph{Proceedings of the 27th
  ACM International Conference on Information and Knowledge Management}}. ACM,
  \bibinfo{pages}{2243--2251}.
\newblock


\bibitem[\protect\citeauthoryear{Paparrizos, Cambazoglu, and Gionis}{Paparrizos
  et~al\mbox{.}}{2011}]%
        {paparrizos2011machine}
\bibfield{author}{\bibinfo{person}{Ioannis Paparrizos},
  \bibinfo{person}{B~Barla Cambazoglu}, {and} \bibinfo{person}{Aristides
  Gionis}.} \bibinfo{year}{2011}\natexlab{}.
\newblock \showarticletitle{Machine learned job recommendation}. In
  \bibinfo{booktitle}{\emph{Proceedings of the fifth ACM Conference on
  Recommender Systems}}. ACM, \bibinfo{pages}{325--328}.
\newblock


\bibitem[\protect\citeauthoryear{Pei, Yang, Cui, Lin, Sun, Jiang, Ou, and
  Zhang}{Pei et~al\mbox{.}}{2019}]%
        {pei2019value}
\bibfield{author}{\bibinfo{person}{Changhua Pei}, \bibinfo{person}{Xinru Yang},
  \bibinfo{person}{Qing Cui}, \bibinfo{person}{Xiao Lin}, \bibinfo{person}{Fei
  Sun}, \bibinfo{person}{Peng Jiang}, \bibinfo{person}{Wenwu Ou}, {and}
  \bibinfo{person}{Yongfeng Zhang}.} \bibinfo{year}{2019}\natexlab{}.
\newblock \showarticletitle{Value-aware Recommendation based on Reinforced
  Profit Maximization in E-commerce Systems}.
\newblock \bibinfo{journal}{\emph{arXiv preprint arXiv:1902.00851}}
  (\bibinfo{year}{2019}).
\newblock


\bibitem[\protect\citeauthoryear{Pizzato, Rej, Chung, Koprinska, and
  Kay}{Pizzato et~al\mbox{.}}{2010a}]%
        {pizzato2010recon}
\bibfield{author}{\bibinfo{person}{Luiz Pizzato}, \bibinfo{person}{Tomek Rej},
  \bibinfo{person}{Thomas Chung}, \bibinfo{person}{Irena Koprinska}, {and}
  \bibinfo{person}{Judy Kay}.} \bibinfo{year}{2010}\natexlab{a}.
\newblock \showarticletitle{RECON: a reciprocal recommender for online dating}.
  In \bibinfo{booktitle}{\emph{Proceedings of the fourth ACM conference on
  Recommender systems}}. ACM, \bibinfo{pages}{207--214}.
\newblock


\bibitem[\protect\citeauthoryear{Pizzato, Rej, Chung, Koprinska, and
  Kay}{Pizzato et~al\mbox{.}}{2010b}]%
        {reciprocal}
\bibfield{author}{\bibinfo{person}{Luiz Pizzato}, \bibinfo{person}{Tomek Rej},
  \bibinfo{person}{Thomas Chung}, \bibinfo{person}{Irena Koprinska}, {and}
  \bibinfo{person}{Judy Kay}.} \bibinfo{year}{2010}\natexlab{b}.
\newblock \showarticletitle{RECON: a reciprocal recommender for online dating}.
  In \bibinfo{booktitle}{\emph{Proceedings of the fourth ACM conference on
  Recommender systems}}. ACM, \bibinfo{pages}{207--214}.
\newblock


\bibitem[\protect\citeauthoryear{Szpektor, Gionis, and Maarek}{Szpektor
  et~al\mbox{.}}{2011}]%
        {szpektor2011improving}
\bibfield{author}{\bibinfo{person}{Idan Szpektor}, \bibinfo{person}{Aristides
  Gionis}, {and} \bibinfo{person}{Yoelle Maarek}.}
  \bibinfo{year}{2011}\natexlab{}.
\newblock \showarticletitle{Improving recommendation for long-tail queries via
  templates}. In \bibinfo{booktitle}{\emph{Proceedings of the 20th
  international conference on World wide web}}. ACM, \bibinfo{pages}{47--56}.
\newblock


\bibitem[\protect\citeauthoryear{Xia, Liu, Sun, and Chen}{Xia
  et~al\mbox{.}}{2015}]%
        {reciprocaldating}
\bibfield{author}{\bibinfo{person}{Peng Xia}, \bibinfo{person}{Benyuan Liu},
  \bibinfo{person}{Yizhou Sun}, {and} \bibinfo{person}{Cindy Chen}.}
  \bibinfo{year}{2015}\natexlab{}.
\newblock \showarticletitle{Reciprocal Recommendation System for Online
  Dating}. In \bibinfo{booktitle}{\emph{Proceedings of the 2015 IEEE/ACM
  International Conference on Advances in Social Networks Analysis and Mining
  2015}}. ACM, \bibinfo{pages}{234--241}.
\newblock


\bibitem[\protect\citeauthoryear{Xiao, Min, Yongfeng, Zhaoquan, Yiqun, and
  Shaoping}{Xiao et~al\mbox{.}}{2017}]%
        {xiao2017fairness}
\bibfield{author}{\bibinfo{person}{Lin Xiao}, \bibinfo{person}{Zhang Min},
  \bibinfo{person}{Zhang Yongfeng}, \bibinfo{person}{Gu Zhaoquan},
  \bibinfo{person}{Liu Yiqun}, {and} \bibinfo{person}{Ma Shaoping}.}
  \bibinfo{year}{2017}\natexlab{}.
\newblock \showarticletitle{Fairness-aware group recommendation with
  pareto-efficiency}. In \bibinfo{booktitle}{\emph{Proceedings of the Eleventh
  ACM Conference on Recommender Systems}}. ACM, \bibinfo{pages}{107--115}.
\newblock


\bibitem[\protect\citeauthoryear{Yao and Huang}{Yao and Huang}{2017}]%
        {yao2017beyond}
\bibfield{author}{\bibinfo{person}{Sirui Yao} {and} \bibinfo{person}{Bert
  Huang}.} \bibinfo{year}{2017}\natexlab{}.
\newblock \showarticletitle{Beyond parity: Fairness objectives for
  collaborative filtering}. In \bibinfo{booktitle}{\emph{Advances in Neural
  Information Processing Systems}}. \bibinfo{pages}{2921--2930}.
\newblock


\bibitem[\protect\citeauthoryear{Yu, Liu, and Zhang}{Yu et~al\mbox{.}}{2011}]%
        {yu2011reciprocal}
\bibfield{author}{\bibinfo{person}{Hongtao Yu}, \bibinfo{person}{Chaoran Liu},
  {and} \bibinfo{person}{Fuzhi Zhang}.} \bibinfo{year}{2011}\natexlab{}.
\newblock \showarticletitle{Reciprocal recommendation algorithm for the field
  of recruitment}.
\newblock \bibinfo{journal}{\emph{JOURNAL OF INFORMATION \&COMPUTATIONAL
  SCIENCE}} \bibinfo{volume}{8}, \bibinfo{number}{16} (\bibinfo{year}{2011}),
  \bibinfo{pages}{4061--4068}.
\newblock


\bibitem[\protect\citeauthoryear{Zhang, Yuan, and Wang}{Zhang
  et~al\mbox{.}}{2014}]%
        {zhang2014optimal}
\bibfield{author}{\bibinfo{person}{Weinan Zhang}, \bibinfo{person}{Shuai Yuan},
  {and} \bibinfo{person}{Jun Wang}.} \bibinfo{year}{2014}\natexlab{}.
\newblock \showarticletitle{Optimal real-time bidding for display advertising}.
  In \bibinfo{booktitle}{\emph{Proceedings of the 20th ACM SIGKDD international
  conference on Knowledge discovery and data mining}}. ACM,
  \bibinfo{pages}{1077--1086}.
\newblock


\bibitem[\protect\citeauthoryear{Zheng, Ghane, and Sabouri}{Zheng
  et~al\mbox{.}}{2019}]%
        {zheng2019personalized}
\bibfield{author}{\bibinfo{person}{Yong Zheng}, \bibinfo{person}{Nastaran
  Ghane}, {and} \bibinfo{person}{Milad Sabouri}.}
  \bibinfo{year}{2019}\natexlab{}.
\newblock \showarticletitle{Personalized Educational Learning with
  Multi-Stakeholder Optimizations}. In \bibinfo{booktitle}{\emph{Adjunct
  Proceedings of the ACM conference on User Modelling, Adaptation and
  Personalization. ACM}}.
\newblock


\end{thebibliography}

\end{document}